\documentclass[]{spie}  

\usepackage{amsmath,amsfonts,amssymb}
\usepackage{graphicx}
\usepackage[colorlinks=true, allcolors=blue]{hyperref}
\usepackage{stackengine}
\usepackage{subcaption}
\usepackage{float}

\usepackage{setspace}
\usepackage{tocloft}
\usepackage{multirow}
\usepackage{lineno}

\usepackage{multirow}

\usepackage{lineno}

\title{Design and validation of a cold load for characterization of CMB-S4 detectors}

\author[a,*]{Cesiley L.~King}
\author[a]{Ian Gullett}
\author[b,c,d]{Adam J.~Anderson}
\author[b,c,d]{Bradford A.~Benson}
\author[a]{Rick Bihary}
\author[a]{Haichen Fan}
\author[a]{Johanna M.~Nagy}
\author[b]{Hogan Nguyen}
\author[a]{John E.~Ruhl}
\author[b]{Sara M.~Simon}

\affil[a]{Physics Department, Case Western Reserve University, 10900 Euclid Ave., Cleveland, OH, 44106, USA}
\affil[b]{Fermi National Accelerator Laboratory, Batavia, IL 60510, USA}
\affil[c]{Department of Astronomy and Astrophysics, University of Chicago, 5640 S. Ellis Ave., Chicago, IL 60637, USA}
\affil[d]{Kavli Institute for Cosmological Physics, University of Chicago, 5640 S. Ellis Ave., Chicago, IL 60637, USA}

\cftpagenumbersoff{figure}
\cftpagenumbersoff{table} 
\begin{document} 
\maketitle

\begin{abstract}
We present the design and validation of a variable temperature cryogenic blackbody source, hereinafter called a cold load, that will be used to characterize detectors to be deployed by CMB-S4, the next-generation ground-based cosmic microwave background (CMB) experiment. Although cold loads have been used for detector characterization by previous CMB experiments, this cold load has three novel design features: (1) the ability to operate from the 1~K stage of a dilution refrigerator (DR), (2) a $^3$He gas-gap heat switch to reduce cooling time, and (3) the ability to couple small external optical signals to measure detector optical time constants under low optical loading. The efficacy of this design was validated using a 150~GHz detector array previously deployed by the \textsc{Spider} experiment. Thermal tests showed that the cold load can be heated to temperatures required for characterizing CMB-S4's detectors without significantly impacting the temperatures of other cryogenic stages when mounted to the DR's 1~K stage. Additionally, optical tests demonstrated that external signals can be coupled to a detector array through the cold load without imparting a significant optical load on the detectors, which will enable measurements of the CMB-S4 detectors' optical time constants. 
\end{abstract}


{\noindent \footnotesize\textbf{*}Cesiley L. King,  \url{cxk554@case.edu} }


\section{Introduction}
\label{sect:intro}  
A vast amount of information about the composition and evolution of the Universe is encoded within the temperature and polarization anisotropies of the cosmic microwave background (CMB), which has driven the development of experiments to measure these anisotropies with increased sensitivity. One such experiment is CMB-S4, which will achieve an unprecedented level of sensitivity by deploying $\sim 500,000$ Transition Edge Sensor (TES) bolometers on a combination of large and small aperture telescopes (LATs and SATs, respectively) located at the South Pole and in the Atacama Desert \cite{S4_science_book, S4_ref_design}. These detectors, fabricated in seven different dichroic wafer designs, will operate in frequency bands ranging from 20 to 280~GHz to facilitate the removal of Galactic and extra-Galactic foregrounds from the CMB signal. The detector wafers are packaged into modules along with feedhorn arrays, time-division multiplexed readout electronics, and mechanical support hardware \cite{DRM_SPIE}.  

CMB-S4's detectors must be characterized prior to deployment to verify that they meet the instrument requirements. This characterization includes measuring  detector properties such as optical efficiency, saturation power, stability, and noise. Typically such measurements are made with the detectors viewing a variable-temperature cryogenic blackbody source,\cite{ACT_coldload, Choi_opt_eff, SPT_coldload, Chuss_coldload} subsequently referred to as a cold load. Additionally, the optical time constants of the detectors should be measured prior to deployment. However, the power radiated by optical sources used in these testing scenarios would typically saturate TES detectors designed to observe the faint CMB signals, so optical time constants are often characterized by proxy through measurements of the electrical time constants \cite{time_const, time_const2}.  However, previous experiments such as \textsc{Spider}\cite{Spider} and SAFARI\cite{SAFARI} have used light pipes to couple external optical signals to measure time constants directly.  

This work describes a cold load that is designed to facilitate rapid characterization of CMB-S4's detector parameters, while also enabling optical characterization of time constants under a range of realistic optical loading conditions. This novel design incorporates three novel features: (1) the ability to operate from the 1~K stage of a dilution refrigerator (DR), (2) a $^3$He heat switch to reduce the cooling time, and (3) the ability to couple small external optical signals into the detectors while they view the cold load.

\section{Design}

The range of temperatures over which the cold load is designed to operate is driven by CMB-S4’s detector properties. These vary with observing frequency band and are listed in Table~\ref{tab:params}. Both the low and high edges of the temperature range are calculated from the target detector saturation power ($P_{sat}$) and predicted on-sky optical loading ($P_{opt}$). The optical powers cover a factor of $\sim 100$ in range, but the required load temperatures only cover a factor of $\sim 10$ because the bands all have roughly equivalent fractional bandwidths;  that is, the higher frequency detectors have broader bands.  These values from CMB-S4 illustrate the broad range of applicable detector parameters for which this cold load's design is well-suited.

The lowest required temperature for each band in Table~\ref{tab:params} is set such that the optical power radiated by the blackbody is 10\% of the detector's designed saturation power, enabling a robust extrapolation to the hypothetical zero temperature load case. Several of the lower-frequency bands require minimum temperatures below 4~K, which can be achieved by mounting the cold load to the $\sim$1~K stage of a DR. The highest required blackbody temperature for a given band is prescribed to be the temperature at which the power radiated by the blackbody is equal to the expected on-sky optical power, which enables a robust detector stability test with that optical loading. This creates a design challenge because the cooling capacity of a typical DR's 1~K stage is in the tens of mW range, much lower than the $\sim$1~W of cooling power at 4~K from a pulse tube cooler. However, since the maximum temperatures required by the highest frequency bands ($\geq 227$~GHz) would exceed this cooling capacity, the cold load will be mounted to 4~K when testing these modules. Previous CMB experiments have successfully used cold loads at 4~K and above \cite{ACT_coldload, Choi_opt_eff}, so many of the tests presented here focus on testing  operation from the DR's 1~K stage.

\begin{table}[htb]
\centering
\caption{The range of cold load temperatures required for each CMB-S4 detector type. The dichroic detector wafers for each telescope type --- Chilean LAT (CHLAT), South Pole LAT (SPLAT) and SAT --- have a variety of band centers, target saturation powers ($P_{sat}$), and anticipated optical loads ($P_{opt}$). The minimum cold load temperature was calculated to produce 10\% of the detector's target saturation power, while the maximum is set to match the anticipated on-sky optical load assuming top-hat bands and 65\% module optical efficiency.
 } 
\label{tab:Paper Margins}
\begin{center}       
\begin{tabular}{||c|c|c|c|c|c|c|c|} 
\hline
\  & $f$ (GHz)  & $P_{sat}$ (pW) & $P_{opt}$ (pW)& Cold Load 
$T_{min}$ (K)& Cold Load $T_{max}$ (K) ) \\ 
\hline
\hline
\multirow{7}{3em}{SPLAT}   & 20.0   & 0.4 & 0.1  & 1.3&2.9   \\  

\cline{2-6}

\   & 24.8   & 0.5& 0.1  &  1.4  &3.0\\   

\   & 36.5   & 2.9& 0.9  & 2.7 & 6.7 \\  

\cline{2-6}

\   &91.5   & 3.8& 1.2  & 3.2 & 6.4 \\   

\   & 148.5   & 10.7 & 3.0 &  5.7& 11.3 \\   

\cline{2-6}

\   & 227.0   & 30.2& 8.2  & 10.3 & 20.7 \\   

\   & 285.5   & 43.9& 11.9  &  14.0 & 28.6\\   
\hline
\hline 
\multirow{6}{3em}{CHLAT}   & 24.8   & 0.5 & 0.2  & 1.4 &3.8 \\   

\   & 36.5   & 2.88 & 1.0 & 2.7 & 7.5 \\   

\cline{2-6}

\   &91.5   & 3.8 & 1.4 & 3.2 & 7.3 \\   

\   & 148.5   & 10.7 & 4.2  &  5.7& 14.6 \\  

\cline{2-6}

\   & 227.0   & 30.2 & 12.0&  10.3& 28.1 \\   

\   & 285.5   & 43.9 & 17.4 & 14.0 & 39.3 \\   
\hline
\hline 
\multirow{8}{2em}{SAT}   & 24.8   & 1.1 & 0.5 & 2.5 & 8.3  \\   

\   & 36.5   & 4.81 & 1.9 & 4.0 & 13.4  \\  

\cline{2-6}

\    & 85.0   & 7.5 & 3.0 &  5.9& 18.3  \\   

\    &145.0   & 10.4& 4.2 &  6.5 & 17.8  \\ 

\cline{2-6}

\   & 95.0  & 7.3 & 2.9 &  5.5 & 16.4\\   

\    & 155.0   & 11.1 & 4.4 & 6.7 & 17.9 \\  

\cline{2-6}

\   & 227.0  & 24.1  & 9.7 & 9.0 & 23.5 \\   

\   & 285.5   & 30.5 & 12.2  & 11.2 & 29.3  \\
\hline
\end{tabular}
\end{center}
\label{tab:params}
\end{table}

\begin{figure}[h]
\centering
\includegraphics[width=14cm]{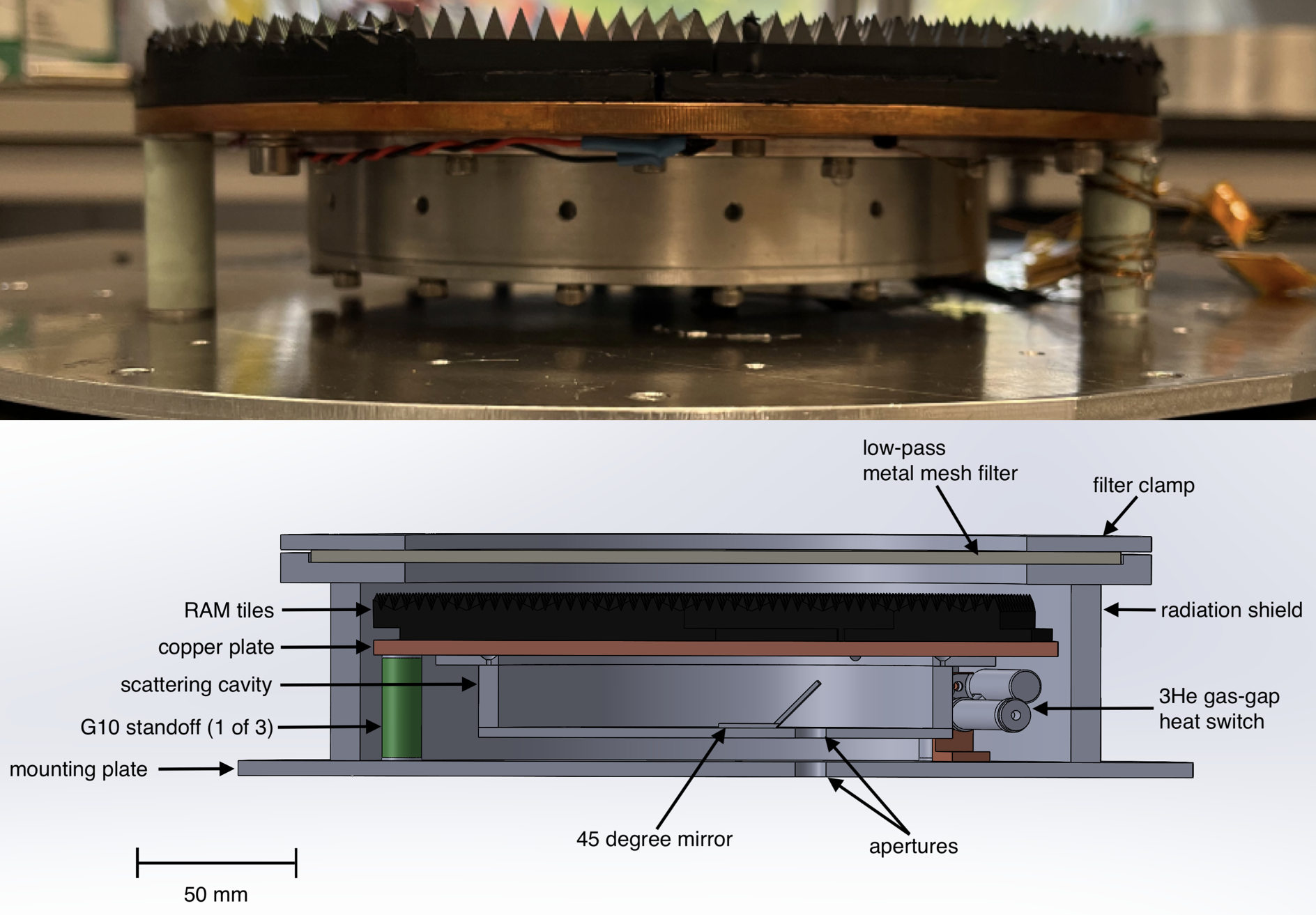}
\caption{Top: Photo of cold load components enclosed within the radiation shield. Bottom: Schematic drawing of the cold load with individual components labeled.}
\label{fig:design}
\end{figure}

Figure \ref{fig:design} shows a photo and schematic drawing of the cold load. It consists of a variable temperature blackbody source that is thermally isolated from a mounting plate and an optical coupling cavity. The physical size of the blackbody load is designed such that it is beam-filling for all detectors on CMB-S4's detector modules, which are $\sim$150~mm-diameter hexagonal structures.\cite{DRM_SPIE}  For CMB-S4's broadest planned beams, which are found in the 90 GHz band of the CHLAT module, the most extreme edge pixels will have $>90$ \% of the beam filled by the cold load in the test cryostats, and this fraction will be higher for all other pixel locations and module types. The cold load is designed to be mounted to the relevant DR temperature stage directly in front of the module, which is mounted to the DR sub-K (100~mK) stage. To minimize the optical power incident on the DR sub-K stage from the heated blackbody, a metal-mesh low-pass filter with a 200~mm-diameter clear aperture is mounted between the detector module and blackbody load\cite{Cardiff_filters}. A 12~icm or lower cutoff frequency is used, depending on the bandpasses of the detector module being tested. Without this filter, the total blackbody optical power emitted through the aperture for a 40~K load is $\sim$5~mW (integrated over all frequencies), which is more than a factor of ten above the cooling power of the DR. Integrating only up to the cutoff frequency of a 12~icm lowpass filter shows that power is reduced to $\sim 16~\mu$W, which is a manageable load even if completely absorbed by the 100~mK DR stage.

The blackbody itself consists of four tessellating THz Radar-Absorbing Material (RAM) tiles from TK Instruments\footnote{\url{https://www.terahertz.co.uk/tk-instruments/products/tesselatingterahertzram}}. Measurements of their performance shared by the manufacturer show less than 25~dB reflectivity in all of CMB-S4's observing bands. \cite{TK_report} The RAM tiles are mounted to a 5~mm thick copper plate with a diameter of 220~mm and the outer edges of the tiles were trimmed such that the tiles could be mounted with no overhang. While each RAM tile is fabricated with alignment pins for joining multiple tiles, these were removed to prevent breakage due to the differential thermal contraction when the tiles are fastened to the copper plate. Each tile is attached to the mounting plate by four retrofitted Trisert inserts\footnote{McMaster-Carr 93722A212} to ensure a good thermal connection to the copper. The inserts are secured with Stycast 2850FT, as this increased their pullout torque from $\sim$1.8~N-m to $\sim$2.6~N-m as measured after ten slow (several hours) thermal cycles to $\sim$80~K.  Electrical power to heat the tiles is provided by three 75~$\Omega$ resistors wired in series and mounted to the backside of the copper plate. The total mass of the variable temperature load, including the RAM tiles and copper plate, is $\sim$2~kg.

The copper plate is mounted via three low-conductivity standoffs to an aluminum base plate that mounts to the cryostat's radiation shield. The standoffs are made of 34~mm long hollow G10 tubes (12.7~mm outer diameter, 0.8~mm wall thickness). These dimensions were chosen based on the thermal conductivity reported in Woodcraft and Gray 2009.\cite{G10} While the standoffs mechanically connect the copper plate and the base plate, their design keeps the two plates sufficiently thermally isolated. There is also a Chase Research Cryogenics $^3$He gas-gap heat switch\footnote{\url{https://www.chasecryogenics.com/?content-block=heat-switches}} with one side mounted to the copper plate and the other to the aluminum base plate to provide a temporary thermal connection to accelerate cooling. The choice of $^3$He to fill the heat switch ensures that there is no possibility of the heat switch leaking superfluid when the cold load is mounted to 1~K. 

A 10~mm diameter aperture in the aluminum base plate is aligned with the aperture of an aluminum optical cavity mounted to the underside of the copper plate. The cavity weighs 0.4~kg, with an inner diameter of 152~mm and a height of 26~mm. Within the optical cavity, there is a 45~degree reflective tab at the entrance of the aperture to prevent any significant fraction of the beam from reflecting immediately back out of the cavity upon reflection from the copper plate. On half of the area of the copper plate and blackbody tiles that coincides with the optical cavity, there is an array of fifty 3~mm diameter holes in a square 13~mm grid pattern to leak the scattered optical signal from the cavity to detectors observing the black side of the cold load. This hole pattern was chosen based on the antenna geometry of a 150~GHz detector array ($\sim$ 8~mm grid) used for the optical measurements described in Section \ref{subsec:optical_meas} previously deployed in the \textsc{Spider} experiment \cite{BKS_dets, Spider_Bmodes}. The holes were drilled into only half of the blackbody in order to probe the effects of coupling vs. distance from a hole.

\section{Performance}
\label{sect:title}

\subsection{Thermal}
To test thermal performance, the cold load was mounted to the 1~K stage of a Bluefors XLD400 DR\footnote{\url{https://bluefors.com/}}. The cryostat was in a ``dark'' configuration, with fully enclosed reflective radiation shields to prevent transmission of light from outside the 4~K space of the cryostat. Thermometers were mounted to the DR's 1~K stage, the detector side of the blackbody RAM tiles, and at the center and edge of the copper tile mounting plate. 

\begin{figure}[h]
\centering
\includegraphics[width=17cm]{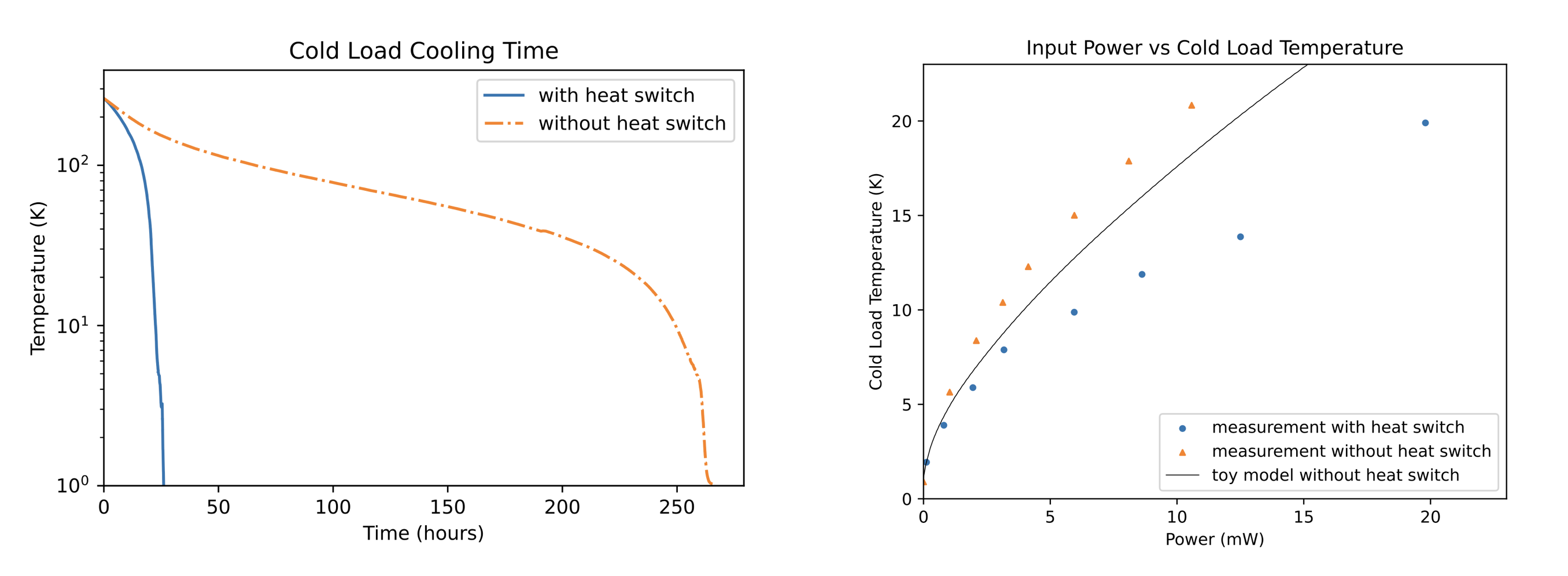}
\caption{Left: Comparison of time required to cool cold load from room temperature to 1~K. Right: Power needed to heat cold load mounted to the 1~K stage of a DR via three G10 standoffs with and without a heat switch installed. A toy model, which uses G10 thermal conductivity from Woodcraft and Gray 2009\cite{G10}, is also shown for reference. 
\label{fig:p_vs_t} }
\end{figure}

A gas-gap heat switch was added to drastically reduce the cooling time from room temperature to 1~K. As shown in Fig.~\ref{fig:p_vs_t} (left), the addition of the heat switch reduced the cooling time from approximately 270 hours to 40 hours. The bare system without the cold load takes approximately 38 hours to cool from room temperature, so 40 hours is a sufficiently fast cooling time. Once the full system is cooled and the cold load is heated for module testing, the inclusion of the heat switch also reduces the cooling time of the cold load from its warmer operating temperatures to 1~K. Without the heat switch, the cold load cools from 20~K to 1~K in $\sim$10~hours, whereas with the heat switch, the cold load cools from 20~K to 1~K in $<$2~hours. Figure~\ref{fig:p_vs_t} (right) shows the electrical power required to heat the blackbody to a given temperature with and without the heat switch installed. Although the ``off" state of the heat switch roughly doubles the power required to reach a given temperature, the load on the 1~K stage remains acceptable for the planned temperature range. 

Measuring the detectors' optical efficiency involves measuring the differences in optical power when the detectors are viewing the blackbody at two (or more) different temperatures.  Any uncertainty in the difference between those temperatures leads to an error in that measurement.  For CMB-S4's module testing, we want to ensure errors smaller than $\sim 5 \%$.  We use Lake Shore Cryotronics 1030 Cernox thermometers\footnote{\url{https://www.lakeshore.com/products/categories/overview/temperature-products/cryogenic-temperature-sensors/cernox}} which we have calibrated to better than $1\%$ in temperature from 1~K to 25~K.  We checked for thermal gradients in the cold load itself by monitoring three thermometers;  one near the center of the copper plate, one near the edge of the copper plate, and one attached to a copper block which was temporarily affixed to the detector side of a blackbody tile.  Heating the blackbody and allowing it to stabilize at six temperatures between 4 and 20~K, we found a mean radial temperature difference of less than 70~mK between the two thermometers on the copper plate as well as between this average and the thermometer on the blackbody tile.  These temperature differences showed no significant trend with load temperature.  If we conservatively adopt 70~mK as the error in measuring blackbody temperature differences, we meet the 5\% measurement criterion by changing the load temperature by 1.4~K or more, which is well within the ranges allowed in Table~\ref{tab:params}. 

\begin{figure}[h]
\centering
\includegraphics[width=13cm]{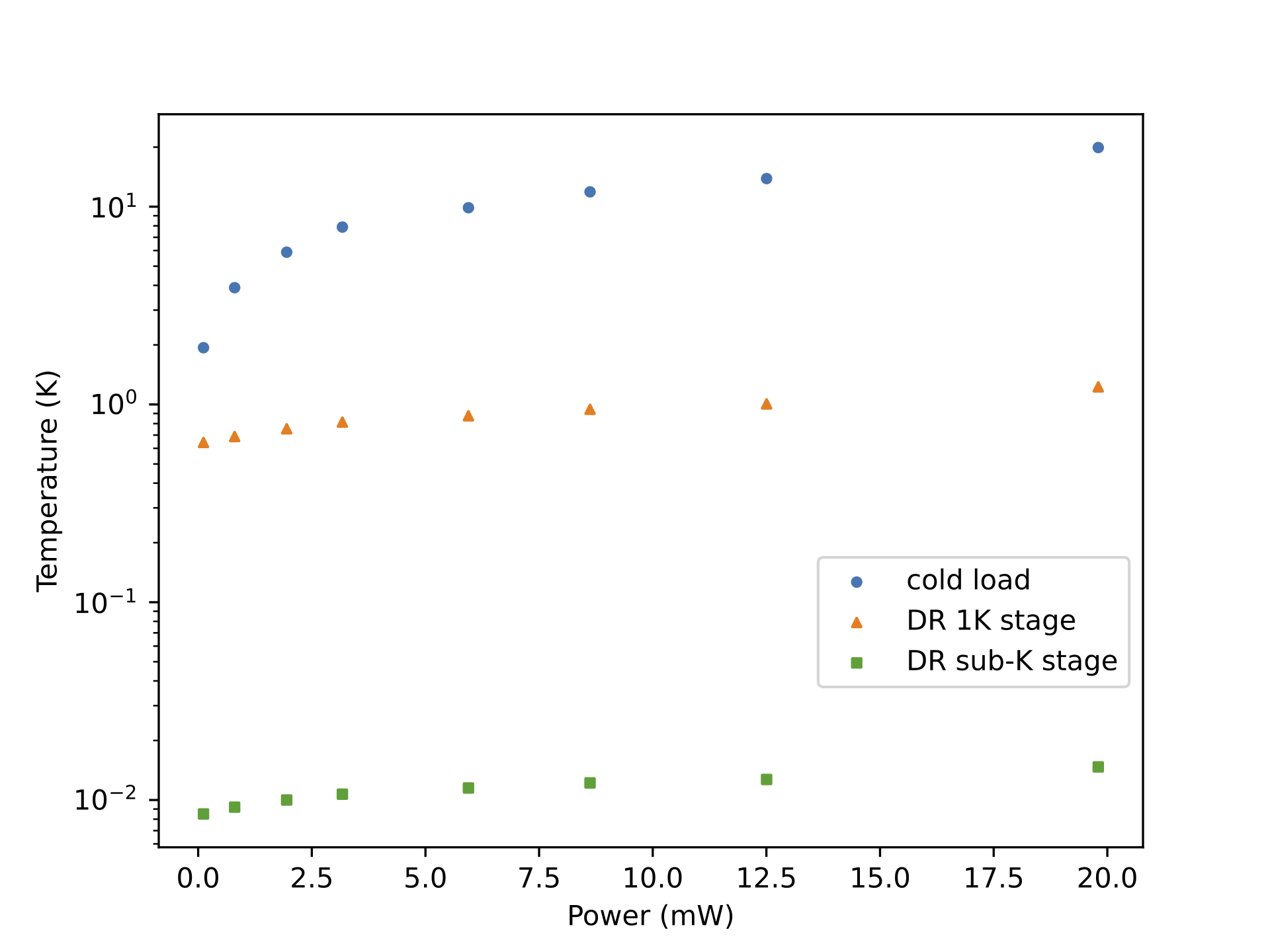}
\caption{Temperatures of the cold load, mounting plate, and 1~K and sub-K stages of the DR for varied cold load heater powers. The cold load and 1~K stage temperatures plotted are the average of the all relevant thermometers.}
\label{fig:temps}
\end{figure}

The 1~K and sub-K stages of the DR do warm slightly when the blackbody is heated, due to the small increases in the incident thermal loading. This is shown in Fig.~\ref{fig:temps}; heating the cold load to 20~K warmed the 1~K stage from 642~mK to 1.2~K, and the sub-K stage from 8.5~mK to 14.7~mK, indicating a load of roughly 10~$\mu$W on the sub-K stage. CMB-S4's detectors are designed to operate from a 100~mK base temperature, where this loading would cause an even smaller temperature rise, roughly 2~mK for this DR, which could be fully compensated for by reducing the heat applied to the sub-K stage to raise it to 100~mK. The source of this extra loading on the sub-K stage is presumably radiative power from the blackbody itself.  Given the area of the cold load, the total radiated blackbody power is roughly $P \approx 300 (T/20\mbox{K})^4$~$\mu$W.  This 300~$\mu$W radiated load at 20~K is similar to the cooling power of the DR at 100~mK.  This motivates the use of the lowpass filter;  at 20~K, the power below its cutoff frequency of 360~GHz is only 7~$\mu$W. 

\subsection{External Optical Coupling}
\label{subsec:optical_meas}
The optical testing capabilities of the cold load were evaluated with a \textsc{Spider} 150~GHz detector array\cite{BKS_dets} mounted to the sub-K stage of an optical Bluefors XLD400 DR. For this testing, the sub-K stage was approximately 300~mK, as this is the operating temperature of the \textsc{Spider} detectors. For these tests, the cold load mounting plate was attached to the DR's 1~K stage.

\begin{figure}[h]
\centering
\includegraphics[width=14cm]{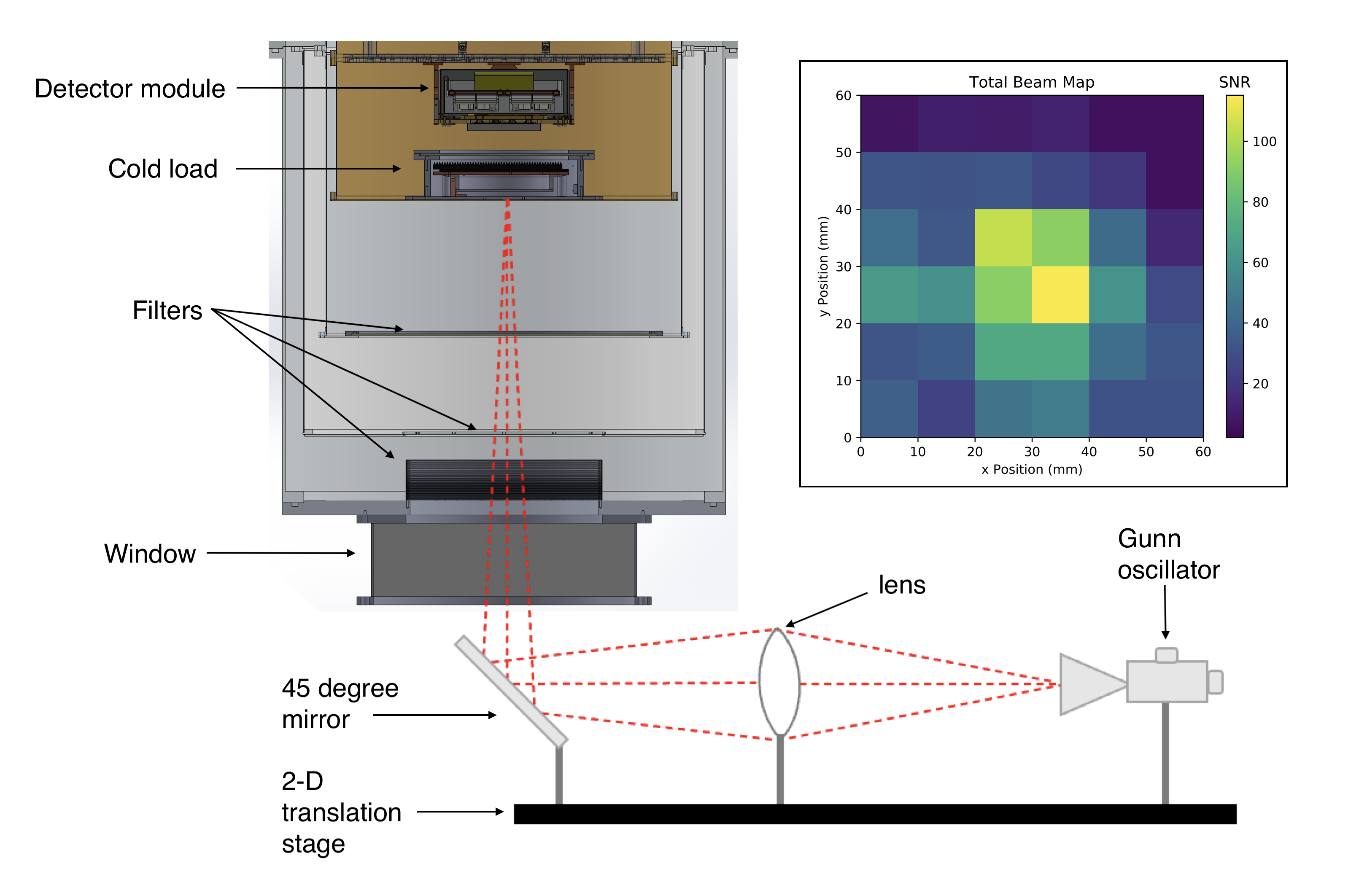}
\caption{Schematic drawing of external optical coupling test setup with relevant components labeled. Red dashed lines indicate the path of the external optical signal. Note that not all components are drawn to scale. Inset: A measured beam map of signal observed by detector array made by varying the source position with the 2-D translation stage. 1-minute detector timestreams were collected in 10~mm~x~10~mm steps over a 60~mm~x~60~mm area, and the chopped signal amplitude on all working detectors were averaged for each map pixel.}
\label{fig:setup}
\end{figure}

Two types of tests were done. The first was to see whether a small chopped optical signal could be 
coupled through the cold load to the detectors to enable time constant characterization under small
optical loading. Apertures in the warmer temperature radiation shields allowed a signal to be coupled from outside the cryostat, as shown in Fig. \ref{fig:setup}, into the cavity behind the 
copper plate.  
A 137~GHz signal from a Gunn oscillator was switched on and off at $\sim$1~Hz to produce a square wave optical power input. This signal was focused by a convex HDPE lens onto a 45~degree mirror, which reflected the signal into the cryostat and ultimately through the cold load apertures and into the cavity mounted to the bottom of the copper plate. Small (3~mm diameter) holes drilled through the copper plate and blackbody tiles allow a small fraction of that power to couple from the cavity to the detectors. 

\begin{figure}[h]
\centering
\includegraphics[width=17cm]{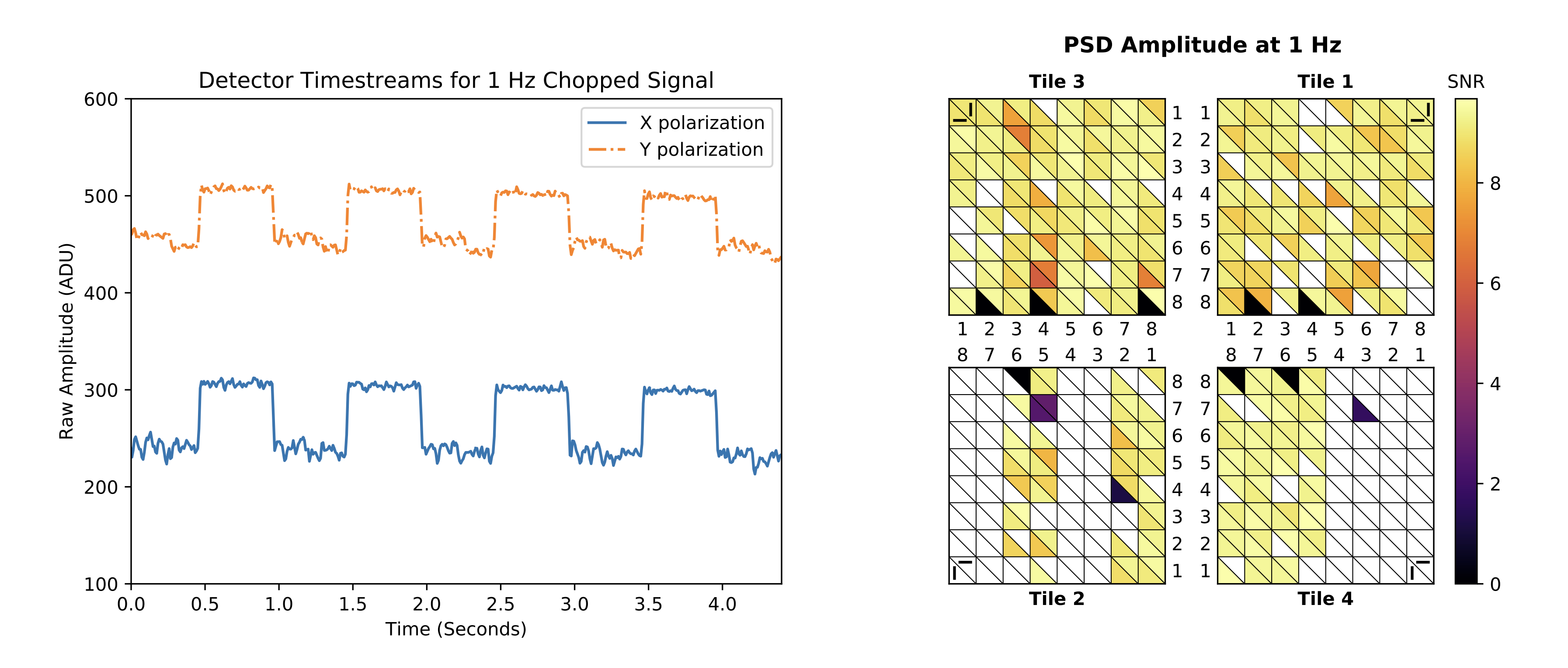}
\caption{Left: Timestreams of two detectors sensitive to orthogonal polarizations X and Y observing a 1~Hz chopped optical signal.
Right: Physical map of the four-tile detector array, color-coded by the signal-to-noise ratio (SNR) of the amplitude of each detector's timestream PSD at 1~Hz. Each small square represents a pixel, and the two triangles within each square represent two detectors sensitive to orthogonal polarizations X and Y. White pixels represent detectors that could not be read out due to poor electrical connections, poorly tuned SQUIDs, or TESs not biased on their transition.  Each of the four square tiles 
is 100~mm across. Tiles 2 and 4 coincide with the half of the cold 
load with 3~mm diameter holes coupling the cavity through 
the blackbody to the detectors.}
\label{fig:timestream}
\end{figure}

Figure~\ref{fig:timestream} (left) shows the timestreams of two detectors sensitive to orthogonal polarizations X and Y observing the 1~Hz chopped signal. Figure~\ref{fig:timestream} (right) shows that the amplitude of this signal is similar for detectors across the entire array, even for those detectors located on the half of the array coinciding with the half of the cold load without holes in the blackbody. This suggests that a more sparse hole pattern could sufficiently illuminate the array in future 
iterations of this cold load design. As expected given the scattering in the cavity, the signal is consistent with being unpolarized at the detectors.

The Gunn oscillator, lens, and mirror were mounted to a 2-D linear translation stage to vary the illumination of the cold load and create ``beam maps" of the coupling through the aperture in the bottom of the cavity. An area of 70~mm~x~70~mm was sampled in 10~mm~x~10~mm steps, and the average signal of all the functioning detectors at each step was used to create the composite beam map shown in the inset in Fig.~\ref{fig:setup}. This beam map displays a distinct central peak, indicating that the signal was being coupled through the cold load cavity rather than through other, more indirect paths to the detectors. 

\begin{figure}[h]
\centering
\includegraphics[width=17cm]{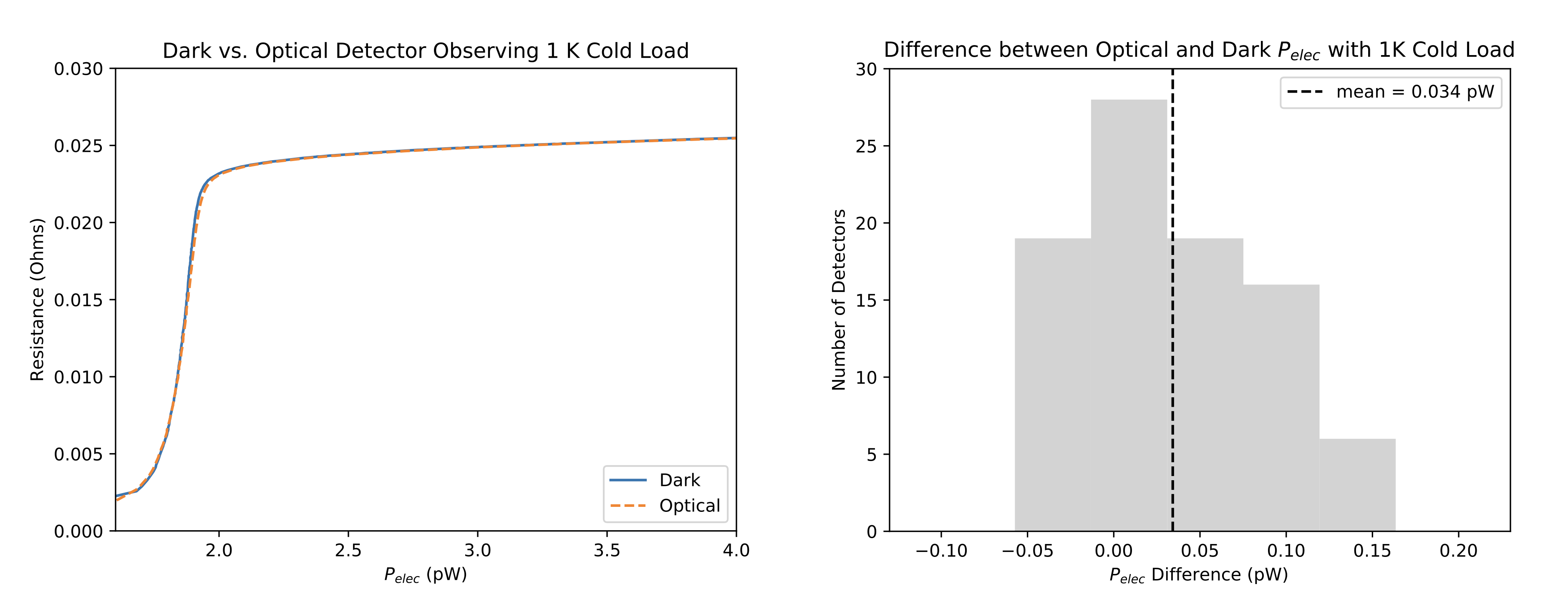}
\caption{Left: Measurements of a detector's resistance as a function of electrical power ($P_{elec}$) observing a 1~K cold load in optical (observing the 1~K cold load) and dark (with the detectors blocked off at 300~mK) test scenarios. The curves look nearly identical, which indicates the additional optical load on the detectors in the optical configuration is very small. Right: Histogram of the difference between the applied $P_{elec}$ required to reach 60\% of $R_n$ in the optical and dark cases, for all detectors. The mean value is indicated by the black dashed line.}
\label{fig:Psat_comp}
\end{figure}

The second test was to check whether the presence of the cavity, with its bottom-side aperture viewing the warmer stages of the cryostat and ultimately the room, generated any significant optical power above that expected from the cold load blackbody tiles.  The diameter and spacing of the holes in the copper plate
and blackbody tiles is such that $\sim 4$\% of the tile area is covered by those holes, suggesting the potential for as much as a 12~K Rayleigh--Jeans load on the detectors if all that optical throughput were coupled to 300~K. The cavity ultimately absorbs much of that throughput.  
To test how much additional optical power this path contributes to the detector loading, we compared 
detector electrical saturation powers measured while observing the 1~K cold load (with its holes and 
apertures open to the room) with those taken with the detectors in a ``fully dark" condition, blanked off at 300~mK. 
Each detector's resistance was measured as a function of bias power in both setups, and the $P_{sat}$ was defined as 60\% of the the normal resistance ($R_n$). Figure \ref{fig:Psat_comp} (left) compares these dark and optical measurements for one detector. Figure \ref{fig:Psat_comp} (right) shows that the mean difference between the dark and optical $P_{sat}$ was 0.034~pW ($\sigma$=0.052). This small difference in $P_{sat}$, less than 1\% of CMB-S4's 150~GHz bands' design saturation powers, indicates that using the cold load in the optical configuration did not contribute a significant amount of optical loading on the detectors. This confirms that cold load is able to provide the intended loading on the detectors in its optical configuration, enabling measurements of detector optical efficiencies and optical time constants in the same cool down.

\section{Conclusion}
This work presents the design and validation of a prototype cold load for CMB-S4 detector module characterization. This cold load is able to operate from the 1~K stage of a DR to reach the low temperatures needed to test CMB-S4's low frequency detector modules. A $^3$He gas-gap heat switch to reduce cooling time from room temperature to allow for an efficient testing schedule. Small external optical signals can be coupled to the detector modules through the cold load to allow optical time constant measurements under representative optical loading conditions. The tests described here demonstrate that this cold load design provides adequate thermal isolation from the DR while maintaining sufficiently low thermal gradients across the blackbody. The signals coupled through the optical cavity are clearly visible across a detector array of similar size to CMB-S4's while maintaining the ability to perform ``dark" characterization measurements on the detectors.

The novel features of this cold load will enable the characterization of all of CMB-S4's various detector module types, and the tests described here will be repeated as CMB-S4's prototype detector modules become available.  Several copies of this cold load design are being fabricated for CMB-S4's single-module test beds. Future work will also include modifying the design for CMB-S4's high throughput test cryostats, each intended to simultaneously characterize seven detector modules and sustain a rapid testing cadence. As design of the high throughput test cryostats is optimized, the impact of the cold load design is being considered to understand the trade offs in the anticipated thermal loads and cooling time.

\subsection*{Disclosures}
The authors have no relevant conflicts of interest to disclose. This work has also been published as conference proceedings for SPIE Astronomical Telescopes + Instrumentation 2024. \cite{cold_load_proceedings}

\subsection* {Code, Data, and Materials Availability} 
The data that support the findings of this paper are not publicly available. They can be requested from the author at cxk554@case.edu.

\subsection* {Acknowledgments}
Work at CWRU was supported by DOE HEP award DE-SC0009946 and NSF award 2240374. This manuscript has been coauthored by Fermi Research Alliance, LLC under Contract No. DE-AC02-07CH11359 with the U.S. Department of Energy, Office of Science, Office of High Energy Physics. The authors would like to thank James Cornelison and Matthew Petroff for serving as the CMB-S4 collaboration's internal reviewers.


\bibliography{report}   

\begin{thebibliography}{10}

\bibitem{S4_science_book}
{Abazajian}, K.~N., {Adshead}, P., {Ahmed}, Z., {Allen}, S.~W., {Alonso}, D., {Arnold}, K.~S., {Baccigalupi}, C., {Bartlett}, J.~G., {Battaglia}, N., {Benson}, B.~A., {Bischoff}, C.~A., {Borrill}, J., {Buza}, V., {Calabrese}, E., {Caldwell}, R., {Carlstrom}, J.~E., {Chang}, C.~L., {Crawford}, T.~M., {Cyr-Racine}, F.-Y., {De Bernardis}, F., {de Haan}, T., {di Serego Alighieri}, S., {Dunkley}, J., {Dvorkin}, C., {Errard}, J., {Fabbian}, G., {Feeney}, S., {Ferraro}, S., {Filippini}, J.~P., {Flauger}, R., {Fuller}, G.~M., {Gluscevic}, V., {Green}, D., {Grin}, D., {Grohs}, E., {Henning}, J.~W., {Hill}, J.~C., {Hlozek}, R., {Holder}, G., {Holzapfel}, W., {Hu}, W., {Huffenberger}, K.~M., {Keskitalo}, R., {Knox}, L., {Kosowsky}, A., {Kovac}, J., {Kovetz}, E.~D., {Kuo}, C.-L., {Kusaka}, A., {Le Jeune}, M., {Lee}, A.~T., {Lilley}, M., {Loverde}, M., {Madhavacheril}, M.~S., {Mantz}, A., {Marsh}, D. J.~E., {McMahon}, J., {Meerburg}, P.~D., {Meyers}, J., {Miller}, A.~D., {Munoz}, J.~B., {Nguyen}, H.~N., {Niemack}, M.~D.,
  {Peloso}, M., {Peloton}, J., {Pogosian}, L., {Pryke}, C., {Raveri}, M., {Reichardt}, C.~L., {Rocha}, G., {Rotti}, A., {Schaan}, E., {Schmittfull}, M.~M., {Scott}, D., {Sehgal}, N., {Shandera}, S., {Sherwin}, B.~D., {Smith}, T.~L., {Sorbo}, L., {Starkman}, G.~D., {Story}, K.~T., {van Engelen}, A., {Vieira}, J.~D., {Watson}, S., {Whitehorn}, N., and {Kimmy Wu}, W.~L., ``{CMB-S4 Science Book, First Edition},'' {\em arXiv e-prints} ,  arXiv:1610.02743 (Oct 2016).
\newblock [doi:10.48550/arXiv.1610.02743].

\bibitem{S4_ref_design}
{Abazajian}, K., {Addison}, G., {Adshead}, P., {Ahmed}, Z., {Allen}, S.~W., {Alonso}, D., {Alvarez}, M., {Anderson}, A., {Arnold}, K.~S., {Baccigalupi}, C., {Bailey}, K., {Barkats}, D., {Barron}, D., {Barry}, P.~S., {Bartlett}, J.~G., {Basu Thakur}, R., {Battaglia}, N., {Baxter}, E., {Bean}, R., {Bebek}, C., {Bender}, A.~N., {Benson}, B.~A., {Berger}, E., {Bhimani}, S., {Bischoff}, C.~A., {Bleem}, L., {Bocquet}, S., {Boddy}, K., {Bonato}, M., {Bond}, J.~R., {Borrill}, J., {Bouchet}, F.~R., {Brown}, M.~L., {Bryan}, S., {Burkhart}, B., {Buza}, V., {Byrum}, K., {Calabrese}, E., {Calafut}, V., {Caldwell}, R., {Carlstrom}, J.~E., {Carron}, J., {Cecil}, T., {Challinor}, A., {Chang}, C.~L., {Chinone}, Y., {Cho}, H.-M.~S., {Cooray}, A., {Crawford}, T.~M., {Crites}, A., {Cukierman}, A., {Cyr-Racine}, F.-Y., {de Haan}, T., {de Zotti}, G., {Delabrouille}, J., {Demarteau}, M., {Devlin}, M., {Di Valentino}, E., {Dobbs}, M., {Duff}, S., {Duivenvoorden}, A., {Dvorkin}, C., {Edwards}, W., {Eimer}, J., {Errard}, J.,
  {Essinger-Hileman}, T., {Fabbian}, G., {Feng}, C., {Ferraro}, S., {Filippini}, J.~P., {Flauger}, R., {Flaugher}, B., {Fraisse}, A.~A., {Frolov}, A., {Galitzki}, N., {Galli}, S., {Ganga}, K., {Gerbino}, M., {Gilchriese}, M., {Gluscevic}, V., {Green}, D., {Grin}, D., {Grohs}, E., {Gualtieri}, R., {Guarino}, V., {Gudmundsson}, J.~E., {Habib}, S., {Haller}, G., {Halpern}, M., {Halverson}, N.~W., {Hanany}, S., {Harrington}, K., {Hasegawa}, M., {Hasselfield}, M., {Hazumi}, M., {Heitmann}, K., {Henderson}, S., {Henning}, J.~W., {Hill}, J.~C., {Hlozek}, R., {Holder}, G., {Holzapfel}, W., {Hubmayr}, J., {Huffenberger}, K.~M., {Huffer}, M., {Hui}, H., {Irwin}, K., {Johnson}, B.~R., {Johnstone}, D., {Jones}, W.~C., {Karkare}, K., {Katayama}, N., {Kerby}, J., {Kernovsky}, S., {Keskitalo}, R., {Kisner}, T., {Knox}, L., {Kosowsky}, A., {Kovac}, J., {Kovetz}, E.~D., {Kuhlmann}, S., {Kuo}, C.-l., {Kurita}, N., {Kusaka}, A., {Lahteenmaki}, A., {Lawrence}, C.~R., {Lee}, A.~T., {Lewis}, A., {Li}, D., {Linder}, E., {Loverde},
  M., {Lowitz}, A., {Madhavacheril}, M.~S., {Mantz}, A., {Matsuda}, F., {Mauskopf}, P., {McMahon}, J., {McQuinn}, M., {Meerburg}, P.~D., {Melin}, J.-B., {Meyers}, J., {Millea}, M., {Mohr}, J., {Moncelsi}, L., {Mroczkowski}, T., {Mukherjee}, S., {M{\"u}nchmeyer}, M., {Nagai}, D., {Nagy}, J., {Namikawa}, T., {Nati}, F., {Natoli}, T., {Negrello}, M., {Newburgh}, L., {Niemack}, M.~D., {Nishino}, H., {Nordby}, M., {Novosad}, V., {O'Connor}, P., {Obied}, G., {Padin}, S., {Pandey}, S., {Partridge}, B., {Pierpaoli}, E., {Pogosian}, L., {Pryke}, C., {Puglisi}, G., {Racine}, B., {Raghunathan}, S., {Rahlin}, A., {Rajagopalan}, S., {Raveri}, M., {Reichanadter}, M., {Reichardt}, C.~L., {Remazeilles}, M., {Rocha}, G., {Roe}, N.~A., {Roy}, A., {Ruhl}, J., {Salatino}, M., {Saliwanchik}, B., {Schaan}, E., {Schillaci}, A., {Schmittfull}, M.~M., {Scott}, D., {Sehgal}, N., {Shandera}, S., {Sheehy}, C., {Sherwin}, B.~D., {Shirokoff}, E., {Simon}, S.~M., {Slosar}, A., {Somerville}, R., {Spergel}, D., {Staggs}, S.~T., {Stark}, A.,
  {Stompor}, R., {Story}, K.~T., {Stoughton}, C., {Suzuki}, A., {Tajima}, O., {Teply}, G.~P., {Thompson}, K., {Timbie}, P., {Tomasi}, M., {Treu}, J.~I., {Tristram}, M., {Tucker}, G., {Umilt{\`a}}, C., {van Engelen}, A., {Vieira}, J.~D., {Vieregg}, A.~G., {Vogelsberger}, M., {Wang}, G., {Watson}, S., {White}, M., {Whitehorn}, N., {Wollack}, E.~J., {Kimmy Wu}, W.~L., {Xu}, Z., {Yasini}, S., {Yeck}, J., {Yoon}, K.~W., {Young}, E., and {Zonca}, A., ``{CMB-S4 Science Case, Reference Design, and Project Plan},'' {\em arXiv e-prints} ,  arXiv:1907.04473 (Jul 2019).
\newblock [doi:10.48550/arXiv.1907.04473].

\bibitem{DRM_SPIE}
{Barron}, D.~R., {Ahmed}, Z., {Aguilar}, J., {Anderson}, A.~J., {Baker}, C.~F., {Barry}, P.~S., {Beall}, J.~A., {Bender}, A.~N., {Benson}, B.~A., {Besuner}, R.~W., {Cecil}, T.~W., {Chang}, C.~L., {Chapman}, S.~C., {Chesmore}, G.~E., {Derylo}, G., {Doriese}, W.~B., {Duff}, S.~M., {Elleflot}, T., {Filippini}, J.~P., {Flaugher}, B., {Gomez}, J.~G., {Grimes}, P.~K., {Gualtieri}, R., {Gullett}, I., {Haller}, G., {Henderson}, S.~W., {Henke}, D., {Herbst}, R., {Huber}, A.~I., {Hubmayr}, J., {Jonas}, M., {Joseph}, J., {King}, C.~L., {Kovac}, J.~M., {Kubik}, D., {Lisovenko}, M., {McMahon}, J.~J., {Moncelsi}, L., {Nagy}, J.~M., {Osherson}, B., {Reese}, B., {Ruhl}, J.~E., {Sapozhnikov}, L., {Schillaci}, A., {Simon}, S.~M., {Suzuki}, A., {Wang}, G., {Westbrook}, B., {Yefremenko}, V., and {Zhang}, J., ``{Conceptual design of the modular detector and readout system for the CMB-S4 survey experiment},'' {\em Millimeter, Submillimeter, and Far-Infrared Detectors and Instrumentation for Astronomy XI}~{\bf 12190},  121900B
  (Aug. 2022).
\newblock [doi:10.1117/12.2630494].

\bibitem{ACT_coldload}
{Henning}, J.~W., {Appel}, J.~W., {Austermann}, J.~E., {Beall}, J.~A., {Becker}, D., {Bennett}, D.~A., {Bleem}, L.~E., {Benson}, B.~A., {Britton}, J., {Carlstrom}, J.~E., {Chang}, C.~L., {Cho}, H.~M., {Crites}, A.~T., {Essinger-Hileman}, T., {Everett}, W., {George}, E.~M., {Halverson}, N.~W., {Hilton}, G.~C., {Holzapfel}, W.~L., {Hubmayr}, J., {Irwin}, K.~D., {Li}, D., {McMahon}, J., {Mehl}, J., {Meyer}, S.~S., {Moseley}, S., {Nibarger}, J.~P., {Niemack}, M.~D., {Parker}, L.~P., {Shirokoff}, E., {Simon}, S.~M., {Staggs}, S.~T., {Ullom}, J.~N., {U-Yen}, K., {Visnjic}, C., {Wollack}, E., {Yoon}, K.~W., {Young}, E.~Y., and {Zhao}, Y., ``{Optical efficiency of feedhorn-coupled TES polarimeters for next-generation CMB instruments},'' {\em Millimeter, Submillimeter, and Far-Infrared Detectors and Instrumentation for Astronomy V}~{\bf 7741},  774122 (July 2010).
\newblock [doi:10.1117/12.859478].

\bibitem{Choi_opt_eff}
{Choi}, S.~K., {Austermann}, J., {Beall}, J.~A., {Crowley}, K.~T., {Datta}, R., {Duff}, S.~M., {Gallardo}, P.~A., {Ho}, S.~P., {Hubmayr}, J., {Koopman}, B.~J., {Li}, Y., {Nati}, F., {Niemack}, M.~D., {Page}, L.~A., {Salatino}, M., {Simon}, S.~M., {Staggs}, S.~T., {Stevens}, J., {Ullom}, J., and {Wollack}, E.~J., ``{Characterization of the Mid-Frequency Arrays for Advanced ACTPol},'' {\em Journal of Low Temperature Physics}~{\bf 193},  267--275 (Nov. 2018).
\newblock [doi:10.1007/s10909-018-1982-4].

\bibitem{SPT_coldload}
Anderson, A., Ade, P., Ahmed, Z., Avva, J., Barry, P., Thakur, R.~B., Bender, A., Benson, B., Bryant, L., Byrum, K., et~al., ``Performance of al--mn transition-edge sensor bolometers in spt-3g,'' {\em Journal of Low Temperature Physics}~{\bf 199},  320--329 (2020).
\newblock [doi:10.1007/s10909-019-02259-7].

\bibitem{Chuss_coldload}
Chuss, D.~T., Rostem, K., Wollack, E.~J., Berman, L., Colazo, F., DeGeorge, M., Helson, K., and Sagliocca, M., ``A cryogenic thermal source for detector array characterization,'' {\em Review of Scientific Instruments}~{\bf 88},  104501 (Oct. 2017).
\newblock [doi:10.1063/1.4996751].

\bibitem{time_const}
Stevens, J.~R., Cothard, N.~F., Vavagiakis, E.~M., Ali, A., Arnold, K., Austermann, J.~E., Choi, S.~K., Dober, B.~J., Duell, C., Duff, S.~M., et~al., ``Characterization of transition edge sensors for the simons observatory,'' {\em Journal of Low Temperature Physics}~{\bf 199},  672--680 (2020).
\newblock [doi:10.1007/s10909-020-02375-9].

\bibitem{time_const2}
Salatino, M., Pappas, C.~G., Henderson, S.~W., Newburgh, L., Niemack, M.~D., Staggs, S.~T., and Wagoner, K., ``Optimization of advanced actpol transition edge sensor bolometer operation using r (t, i) transition measurements,'' {\em IEEE Transactions on Applied Superconductivity}~{\bf 27}(4),  1--6 (2017).
\newblock [doi:0.1109/TASC.2017.2672687].

\bibitem{Spider}
{SPIDER Collaboration}, {Ade}, A.~R., {Amiri}, M., {Benton}, S.~J., {Bergman}, A.~S., {Bihary}, R., {Bock}, J.~J., {Bond}, J.~R., {Bonetti}, J.~A., {Bryan}, S.~A., {Chiang}, H.~C., {Contaldi}, C.~R., {Dor{\'e}}, O., {Duivenvoorden}, A.~J., {Eriksen}, H.~K., {Farhang}, M., {Filippini}, J.~P., {Fraisse}, A.~A., {Freese}, K., {Galloway}, M., {Gambrel}, A.~E., {Gandilo}, N.~N., {Ganga}, K., {Gualtieri}, R., {Gudmundsson}, J.~E., {Halpern}, M., {Hartley}, J., {Hasselfield}, M., {Hilton}, G., {Holmes}, W., {Hristov}, V.~V., {Huang}, Z., {Irwin}, K.~D., {Jones}, W.~C., {Karakci}, A., {Kuo}, C.~L., {Kermish}, Z.~D., {Leung}, J. S.-Y., {Li}, S., Y., M. D.~S., {Mason}, P.~V., {Megerian}, K., {Moncelsi}, L., {Morford}, T.~A., {Nagy}, J.~M., {Netterfield}, C.~B., {Nolta}, M., {O'Brient}, R., {Osherson}, B., {Padilla}, I.~L., {Racine}, B., {Rahlin}, A.~S., {Reintsema}, C., {Ruhl}, J.~E., {Runyan}, M.~C., {Ruud}, T.~M., {Shariff}, J.~A., {Shaw}, E.~C., {Shiu}, C., {Soler}, J.~D., {Song}, X., {Trangsrud}, A., {Tucker}, C.,
  {Tucker}, R.~S., {Turner}, A.~D., {van der List}, J.~F., {Weber}, A.~C., {Wehus}, I.~K., {Wen}, S., {Wiebe}, D.~V., and {Young}, E.~Y., ``A constraint on primordial b-modes from the first flight of the spider balloon-borne telescope,'' {\em The Astrophysical Journal}~{\bf 927}(2),  174 (2022).

\bibitem{SAFARI}
{Audley}, M.~D., {Lange}, G.~d., {Gao}, J.-R., {Khosropanah}, P., {Mauskopf}, P.~D., {Morozov}, D., {Trappe}, N.~A., {Doherty}, S., and {Withington}, S., ``{Optical characterization of ultra-sensitive TES bolometers for SAFARI},'' {\em Millimeter, Submillimeter, and Far-Infrared Detectors and Instrumentation for Astronomy VII}~{\bf 9153},  91530E (July 2014).
\newblock [doi:110.1117/12.2056037].

\bibitem{Cardiff_filters}
{Tucker}, C.~E. and {Ade}, P. A.~R., ``{Thermal filtering for large aperture cryogenic detector arrays},'' {\em Millimeter and Submillimeter Detectors and Instrumentation for Astronomy III}~{\bf 6275},  62750T (June 2006).
\newblock [doi:10.1117/12.673159].

\bibitem{TK_report}
{Saenz}, E., ``{TK RAM} reflectifity and diffuse scattering test campaign - 18-26 {GHz}.'' \url{https://www.terahertz.co.uk/images/tki/RAM/TK_ESA_RAM_Low.pdf} (2019).

\bibitem{G10}
{Woodcraft}, A.~L. and {Gray}, A., ``{A low temperature thermal conductivity database},'' {\em Low Temperature Detectors LTD-13}~{\bf 1185},  684 (Dec. 2009).
\newblock [doi:10.1063/1.3292433].

\bibitem{BKS_dets}
Ade, P.~A., Aikin, R., Amiri, M., Barkats, D., Benton, S., Bischoff, C.~A., Bock, J., Bonetti, J., Brevik, J., Buder, I., et~al., ``Antenna-coupled tes bolometers used in bicep2, keck array, and spider,'' {\em The Astrophysical Journal}~{\bf 812}(2),  176 (2015).
\newblock [doi:10.1088/0004-637X/812/2/176].

\bibitem{Spider_Bmodes}
{Ade}, P.~A.~R., {Amiri}, M., {Benton}, S.~J., {Bergman}, A.~S., {Bihary}, R., {Bock}, J.~J., {Bond}, J.~R., {Bonetti}, J.~A., {Bryan}, S.~A., {Chiang}, H.~C., {Contaldi}, C.~R., {Dor{\'e}}, O., {Duivenvoorden}, A.~J., {Eriksen}, H.~K., {Farhang}, M., {Filippini}, J.~P., {Fraisse}, A.~A., {Freese}, K., {Galloway}, M., {Gambrel}, A.~E., {Gandilo}, N.~N., {Ganga}, K., {Gualtieri}, R., {Gudmundsson}, J.~E., {Halpern}, M., {Hartley}, J., {Hasselfield}, M., {Hilton}, G., {Holmes}, W., {Hristov}, V.~V., {Huang}, Z., {Irwin}, K.~D., {Jones}, W.~C., {Karakci}, A., {Kuo}, C.~L., {Kermish}, Z.~D., {Leung}, J.~S.~Y., {Li}, S., {Mak}, D.~S.~Y., {Mason}, P.~V., {Megerian}, K., {Moncelsi}, L., {Morford}, T.~A., {Nagy}, J.~M., {Netterfield}, C.~B., {Nolta}, M., {O'Brient}, R., {Osherson}, B., {Padilla}, I.~L., {Racine}, B., {Rahlin}, A.~S., {Reintsema}, C., {Ruhl}, J.~E., {Runyan}, M.~C., {Ruud}, T.~M., {Shariff}, J.~A., {Shaw}, E.~C., {Shiu}, C., {Soler}, J.~D., {Song}, X., {Trangsrud}, A., {Tucker}, C., {Tucker}, R.~S.,
  {Turner}, A.~D., {van der List}, J.~F., {Weber}, A.~C., {Wehus}, I.~K., {Wen}, S., {Wiebe}, D.~V., {Young}, E.~Y., and {Spider Collaboration}, ``{A Constraint on Primordial B-modes from the First Flight of the SPIDER Balloon-borne Telescope},'' {\em The Astrophysical Journal}~{\bf 927},  174 (Mar. 2022).
\newblock [doi:10.3847/1538-4357/ac20df].

\bibitem{cold_load_proceedings}
{King}, C.~L., {Gullett}, I., {Anderson}, A.~J., {Benson}, B.~A., {Bihary}, R., {Fan}, H., {Nagy}, J.~M., {Nguyen}, H., {Ruhl}, J.~E., and {Simon}, S.~M., ``{Design and validation of a cold load for characterization of CMB-S4 detectors},'' {\em Millimeter, Submillimeter, and Far-Infrared Detectors and Instrumentation for Astronomy XII}~{\bf 13102},  1310218 (Aug. 2024).
\newblock [doi:10.1117/12.3018159].

\end{thebibliography}
\bibliographystyle{spiebib}   

\end{document}